\documentclass[aps,prb,twocolumn,superscriptaddress]{revtex4-1}
\usepackage{graphics}
\usepackage{longtable}
\usepackage{epsfig}
\usepackage{amsmath}
\usepackage{float}
\usepackage{color}
\usepackage{bigints}
\usepackage{amsmath}
\usepackage{amsfonts}
\usepackage{amssymb}
\usepackage{amsmath}
\usepackage{amsfonts}
\usepackage{amssymb}
\usepackage[flushleft]{threeparttable}

\usepackage[version=3]{mhchem}

\begin{document}

\title{Impressive computational acceleration by using machine learning for 2-dimensional super-lubricant materials discovery}

\author{Marco \surname{Fronzi}}
\email[Corresponding author\\ E-mail: ]{marco.fronzi@uts.edu.au}
\affiliation{International Research Centre for Renewable Energy, State Key Laboratory of Multiphase Flow in Power Engineering, Xi'an Jiaotong University, Xi'an 710049, Shaanxi, China}
\affiliation{School of Mathematical and Physical Science, University of Technology Sydney, Ultimo 2007, New South Wales 2007, Australia}

\author{Mutaz \surname{Abu Ghazaleh}}
\affiliation{School of Mathematical and Physical Science, University of Technology Sydney, Ultimo 2007, New South Wales 2007, Australia}

\author{Olexandr  \surname{Isayev}}
\affiliation{Laboratory for Molecular Modeling, Division of Chemical Biology and Medicinal Chemistry, UNC Eshelman School of Pharmacy, University of North Carolina, Chapel Hill, NC 27599, USA}

\author{David  A.\surname{Winkler}}
\affiliation{Manufacturing, Commonwealth Scientific and Industrial Research Organization,
Bag 10, Clayton South MDC, Victoria 3169, Australia}
\affiliation{Monash Institute of Pharmaceutical Sciences, Monash University,
381 Royal Parade, Parkville, Victoria 3052, Australia}
\affiliation{Latrobe Institute for Molecular Science, La Trobe University,
Kingsbury Drive, Bundoora, Victoria 3086, Australia}
\affiliation{School of Pharmacy, The University of Nottingham, Nottingham NG7 2RD, UK}

\author{Joe  \surname{Shapter}}
\affiliation{Australian Institute for Bioengineering and Nanotechnology,
The University of Queensland, St Lucia, Brisbane, Queensland 4072,
Australia and College of Science and Engineering, Flinders University,
Bedford Park, Adelaide, South Australia 5042, Australia}

\author{Michael J.  \surname{Ford}}
\email[Corresponding author\\ E-mail: ]{Mike.Ford@uts.edu.au}
\affiliation{School of Mathematical and Physical Science, University of Technology Sydney, Ultimo 2007, New South Wales 2007, Australia}

\date{\today}

\begin{abstract}

The screening of novel materials is an important topic in the field of materials science. Although traditional computational modelling, especially first-principles approaches, is a very useful and accurate tool to predict the properties of novel materials, it still demands extensive and expensive state-of-the-art computational resources. Additionally, they can be often extremely time consuming. We describe a time and resource efficient machine learning approach to create a large dataset of structural properties of van der Waals layered structures. In particular, we focus on the interlayer energy and the elastic constant of layered materials composed of two different 2-dimensional (2D) structures, that are important for novel solid lubricant and super-lubricant materials. We show that machine learning models can recapitulate results of computationally expansive approaches (i.e. density functional theory) with high accuracy.

\end{abstract}

\maketitle


\section{Main}

Solid lubricants are materials that reduce friction and damage of contacting surfaces in relative motion. A good lubricant can be identified by two main properties: shear strength and abrasivity. The dynamics of solid lubrication show that sliding motion is followed by significant ductile shear of the solid lubricant film. Therefore, the solid lubricant must have low shear strength, which occurs in crystalline phase by slip along preferred crystallographic planes.\cite{Sliney1978} On the other hand, abrasivity is a relative property that is a function of the hardness ratio of the lubricant and the lubricated material. Typically, the lubricant particles should be softer than the contact material to avoid abrasions.\cite{Sliney1978} Clearly, thermo-chemical stability in the environment of the application is also essential. This is particularly important for high temperature applications, but is equally important for moderate temperature applications to insure adequate storage stability and to avoid corrosion by atmospheric components such as oxygen and salt spray. According to this description, thermo-chemically stable materials, with a low interlayer energy and low degree of hardness, are good candidates for solid lubricant materials.
A significant drawback for the application of conventional structural materials concerns the strong anisotropic nature of friction in homogeneous and heterogeneous interfaces with respect to their relative orientation. Even when the interface is formed in an incommensurate ultra-low friction configuration, the contact surfaces have a tendency to rotate towards the aligned commensurate configuration during the sliding motion and eventually lock in a high friction state, which corresponds to a higher interaction energy.\cite{Filippov2013}
Low-energy-interaction/high-shear-motion structures can be found in novel 2-dimensional (2D) van der Waals (vdW) layered structures.\cite{Novoselov2016,Geim2013} Van der Waals forces differ from covalent and ionic bonding in that they are caused by correlations in the fluctuating polarizations of nearby particles, resulting in weak, long range forces. The vdW strength of two contacting structures is a key requirement for lubricity/superlubricity behaviour. \cite{Wang2018}
Currently, only a few Van der Waals structures, graphite, boron nitride and molybdenum disulfide, are used as dry lubricants. However, it is possible to theoretically identify 6,138 single layer 2D materials, available  from an online database (https://2dmatpedia.org/), obtained by theoretical exfoliation and elemental substitution  from a large number of inorganic bulk structures included in the  database Materials Project (https://materialsproject.org/).\cite{Jain2013,Zhou2019} By a direct stacking of  2D materials, it is possible to generate 18,834,453 unique novel bilayers hetero-structures ($N_b=N_m(N_m+1)/2$).\cite{Ponomarenko2011,Britnell2012,Haigh2012} Furthermore, previous research has shown that 2D materials stacking processes are self-cleaning, resulting in near ideal hybrid 2D layered structures.\cite{Georgiou2013,Wang2013} Due to the large number of structures and to the nature of the vdW forces that are responsible for the 2D monolayers stacking,  it is likely that within the generated hetero-structures set there will be many with low interaction energies, and eventually a desirable softness and temperature stability to be used in super-lubricant applications. However, due to the extremely large number of possible layer combinations, exhaustive experimental assessment is clearly infeasible. Traditional quantum chemical computational techniques, such density functional theory (DFT),  can accurately predict the properties of such materials and can be used for the discovery of new materials, however, the computational demands of these calculations means the materials assessment process is still very slow. \cite{Goedecker1999,Bowler2012}

Here, we propose a time and resources efficient machine learning (ML) approach that, combined with a limited number of first-principles calculations, is able to calculate and predict the interlayer energy (IE) and the elastic constant of a large number of layered hetero-structures,  expanding  the capabilities of a canonical thoery. We use DFT to predict the desired properties of a relatively small number of 2D layered hybrid  materials. We leverage this smaller pool of results by using them to train  supervised machine learning models. The models can then rapidly and reliably predict these quantities for a large number of structures ($\sim$18M bilayers) within the domain of applicability of the models. We use a Bayesian neural network (BNN) model, which allows us to have a confidence interval for our predictions.\cite{Correa2009,Butler2018}

We calculated the interlayer energies and the elastic constant for two representative subsets of all the possible combinations of two 2D materials, consisting of 282 structures for IE and 226 for elastic constant, which here is fairly approximated,  to the C$_{33}$ value.\cite{Bertolazzi2011,Liu2016}
Members of these subsets had an interlayer energy $E\leq-1.0$ eV/{\AA}$^2$ and a maximum lattice constant mismatch of 2$\%$. 
Due to the nature of the vdW forces, the interlayer energies of 2D hetero-structures depend weakly on the twisting angle between the specific stacking configurations, typically by $\leq$30 meV. Therefore, it is important to point out that any twist angle would not affect our conclusions. This makes the problem of finding low friction structures less difficult from a computational perspective.\cite{Lu2017} Consequently, for simplicity, we set the twist angle between the two monolayers to be 0 degrees. 
 Finally, the  temperature stability of each bilayer was estimated by considering the minimum  value between the two monolayers decomposition energy within each hetero-structure. 
The interlayer energies that we obtain agree well with the available data in the literature, as shown in Table \ref{I-E}. However, our calculated values of C$_{33}$  are significantly smaller than the experimental values, and show deviations from both the values calculated using other vdW approximations (e.g. vdW-DF1, vdW-DF2, vdW-VV10). In general, vdW potentials are less accurate for calculating the elastic constants compared to the interlayer energy. This can be attributed to the limits of the current vdW approximations.\cite{Bjrkman2012,Rego2015,Bjorkman2012b}   Here, we report only a few examples that highlight discrepancies and agreements with other work, however, a wider comparison can be found in the literature, particularly in the work of Bj\"{o}rkman {\em et al.}\cite{Bjrkman2012,Lebegue2010,Liu2011,Zacharia2004,Graziano2012,Wang2017,Rydberg2003}

\begin{figure*}[ht]
\centering
  \begin{tabular}{@{}cccc@{}}
     \includegraphics[width=.9\textwidth]{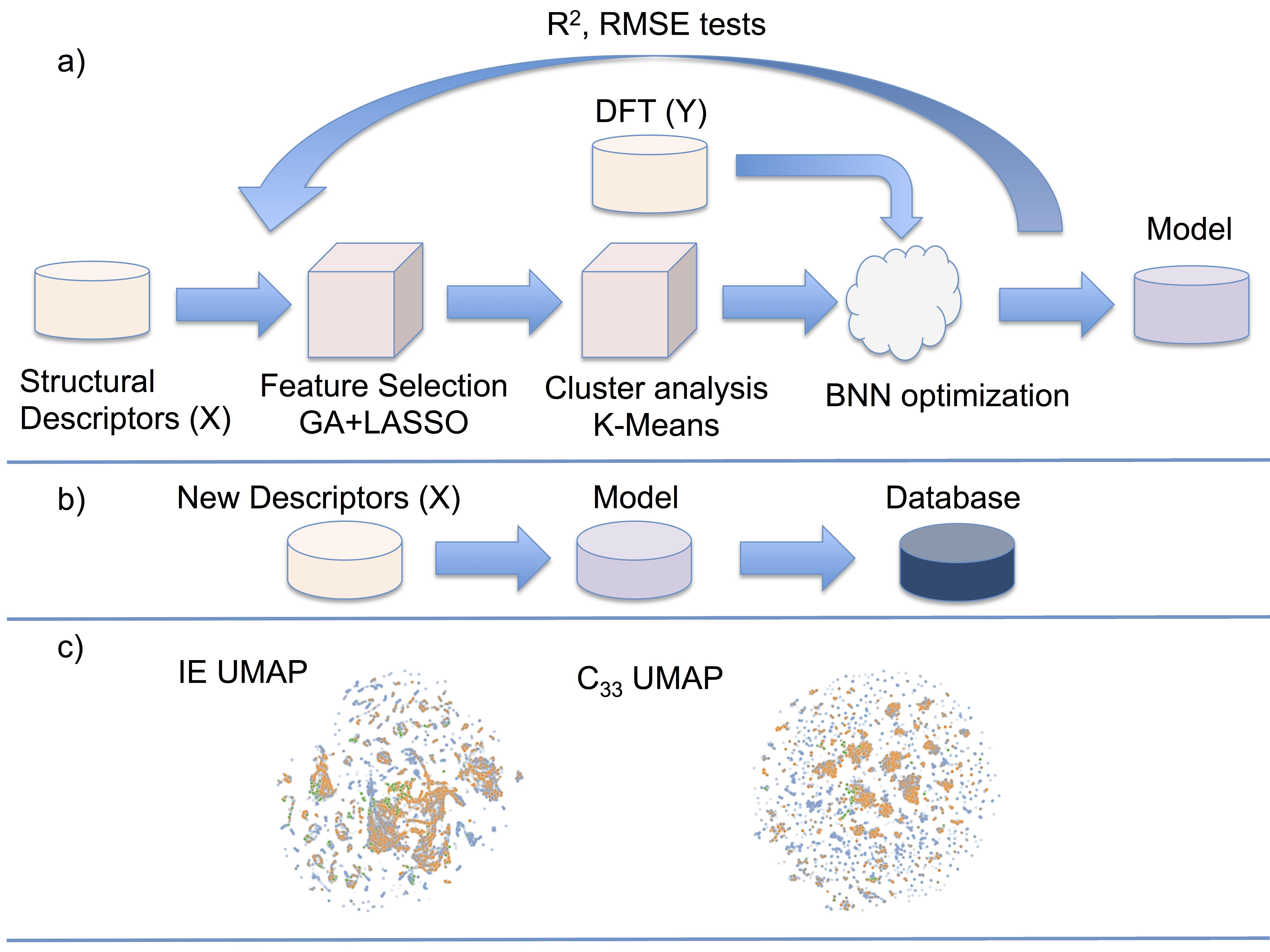} \\
  \end{tabular}
  \caption{(Color online) a) Schematic representation of the workflow used to create our BNN model and b) to extrapolate IE and C$_{33}$ values using a set of descriptors for new structures. c) UMAPs if the of the IE and C$_{33}$, respectively. The blue, orange and green dots represent the total representative bilayers subset, the commensurate unit cells, and the DFT calculations, respectively.}
  \label{fig:pic0}
\end{figure*}

The DFT results were used to train supervised machine learning (ML) models of interlayer energies and elastic constants. The Bayesian neural networks that we employed require a target property ($Y$), calculated by means of DFT,  and a list of descriptors ($X$) (mathematical objects that represent the molecular properties of the materials) for each bilayer. 
We generated descriptors for each monolayer using the method developed by us previousely which have been shown to be useful in previous work.\cite{Isayev2017,Tawfik2019} Descriptors for each bilayer were obtained by adding the values of the descriptors for the two monolayers, as described by Tawfik et al.\cite{Tawfik2019} The algorithm generated 2,764 descriptors for each monolayer. To avoid problems of overfitting due to this large number of descriptors, we selected a small subset of the most relevant features after using a combination of Genetic Algorithm (GA) search and LASSO regression. This reduced the number of descriptors to 42 for the interlayer energy model and 89 for the C$_{33}$ model. The GA and LASSO eliminate irrelevant or low relevance features, which makes the ML models train more quickly, generalize to new data better, and easier to interpret.\cite{Cai2018} The same subset of relevant descriptors is used for predicting the training and test sets, and for generating the properties of the $\sim$18M structure superset.\cite{Isayev2017,Tawfik2019}  The same subset of relevant descriptors is used for predicting the training and test sets, and for generating the properties of the $\sim$18M structure superset.  
The domain of applicability of the model to the large dataset is ensured by the equivalent range of the descriptors values in the large dataset and in the subset used in the training. 
To obtain the train and test set in a way that each subgroup contains representative structures of the total set, we perform a cluster analysis. This will maximize the diversity of structures assigned to training and test sets while ensuring that the test sets are still within the domain of the models.
The subset used in the training is represented in the UMAP in Figure \ref{fig:pic0} c), which shows how the DFT calculations are distributed over the commensurate bilayers, and over the whole dataset obtained by subsampling. The UMAP, gives also an additional information on the quaisi-uniform distribution of the commensurate cells over the whole set. This was estimated to be  $\sim$8.2$\%$ of the total number of bilayers, considering two constrains: the lattice mismatch $L_m\leq2\%$, and the number of atoms in the cell to be $N_a\leq600$. The first constraint ensures a reliable outcome of DFT calculations with respect to a realistic scenario; the second constraint ensures that the DFT calculations can be performed within a reasonable computational time, which, in conventional DFT calculations, scales cubically to the of Kohn$-$Sham orbitals in target systems.\cite{Tamleh2018,Goedecker1999,Bowler2012} 

After BNN optimization, performed using cross-validation subsets, we test the quality of the results by calculating R$^2$, Root Mean Square Error and Mean Absolute Error, as shown in Table \ref{R2IE}. Here, we consider Root Mean Square Error more indicative than R$^2$ as they have been shown to be less dependent on the number of training samples and complexity of the models.\cite{Alexander2015}
Our implementation uses a dropout approach that is used during training for regularization, and during predictions, to obtain a statistical distribution of the response, which provide information on the uncertainty of the values, as described in Section \ref{BNN}.
To further test the quality of our predictions we perform an additional test on a validation set, which was not used during BNN training.

\begin{figure*}[ht]
\centering
  \begin{tabular}{@{}cccc@{}}
     \includegraphics[width=1.0\textwidth]{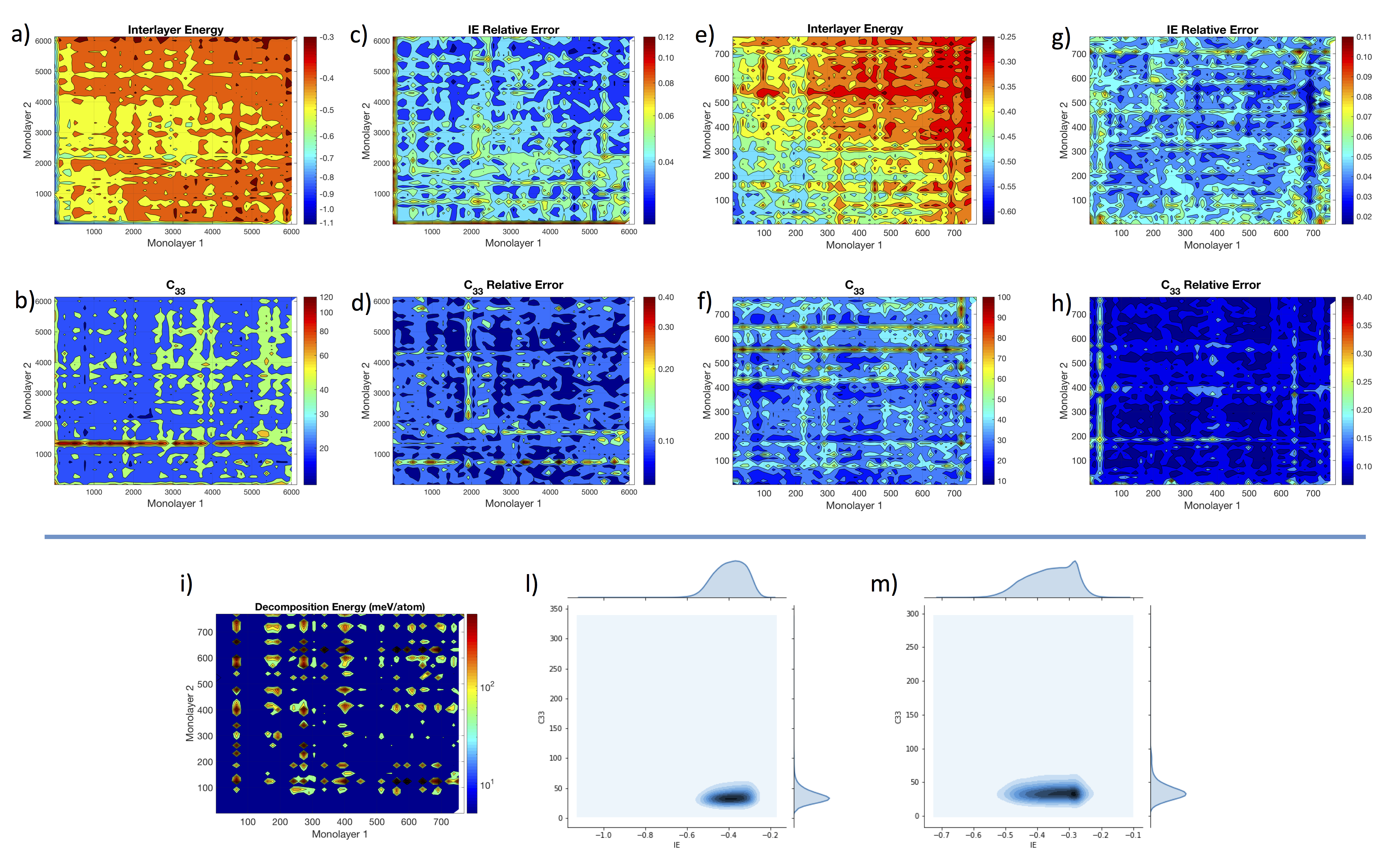} \\

  \end{tabular}
  \caption{(Color online) a) and b) Interlayer Energies and C$_{33}$ of the $\sim$18M bilayer set. c) and d) Relative Error of the Interlayer Energies and C$_{33}$ of the $\sim$18M bilayer set. e) and f) Interlayer Energies and C$_{33}$ of the $\sim$300k bilayer set. g) and h) Relative Error of the Interlayer Energies and C$_{33}$ of the $\sim$300k bilayer set. e) Temperature stability of the $\sim$300k bilayer set. f) and g)  Statistical distribution of IE and C$_{33}$  values in the $\sim$18M and in the  $\sim$300k datasets, respectively.  
Here, Interlayer Energies are expressed in eV/{\AA}$^2$, the C$_{33}$ in GPa and temperature stability in energy per atom (meV/atom).  
  Absolute errors have been calculated as the standard deviation of the response distribution, using a dropout approach with probability 0.1. Detailed information can be found in Section Methods. }
  \label{fig:pic1}
\end{figure*}

Using this model we extrapolated the interlayer energy and C$_{33}$ for 18,834,453  structures and the complete list of values, together with the associated uncertainty, is available online (http://doi.org/10.26195/5dd36650d7e1e). Figures \ref{fig:pic1} a) and c) show the interlayer energy and C$_{33}$ as a heatmap. Here, each axis contains the list of monolayers that forms the bilayers, and the axes are ordered to cluster structures with similar values of the interlayer energy. The order of the monolayer in the axes is the same in the interlayer energy and C$_{33}$ plots, therefore the sparse clustering in C$_{33}$ map, suggest a weak correlation between the two properties. Figures \ref{fig:pic1} b) and d) show the the relative error calculated for the interlayer energy and C$_{33}$. The map suggests that our BNN is particularly inaccurate in predicting hard materials, indicated by the coloured areas, which represent a small fraction ($\leq1\%$) of the overall results.
However, the remainder of the map suggests a reliable prediction with an average accuracy of $\sim$4$\%$ for the interlayer energy and $\sim$11$\%$ for the C$_{33}$.

Considering the absolute value of the IE, we obtain a Pearson value 0.06 in ML and 0.05 in DFT, and Spearman value of 0.09 in ML and 0.08 in DFT, indicating an extremely weak relationship between the two properties. However, the same coefficients are higher in the subset of homo-bilayer, where the Pearson and Spearman values are 0.40 and 0.55. 
Although there is no correlation between the two quantities, we calculated that 90$\%$ of the bilayers has an IE  between $-$0.51 eV and $-$0.28 eV and a C$_{33}$ between 19.44 GPa and 63.44 GPa, as shown in Figure \ref{fig:pic1} l) and m). Due to the very large size of the dataset considered here, we can generalize this conclusion to all the possible van der Waals hetero-structures formed by assembling any combination of two 2D material.

\begin{table*}[ht]
\caption{\label{I-E} Interlayer energies (IE)  (in eV/{\AA}$^2$) and C$_{33}$  (in GPa)  measured and calculated by DFT (using vdW-VV10, vdW-DF1, vdW-DF2) as reported in the literature, and calculated in the present work using DFT (vdW-TS). }
\centering
\begin{tabular}{p{0.20\linewidth}p{0.2\linewidth}p{0.2\linewidth}p{0.2\linewidth}p{0.20\linewidth}}
\hline
\hline
2D bilayer  & Calculated IE  &  Measured Value & Calculated C$_{33}$ & Measured Value\\
\hline
Graphene & $-$0.39\cite{Bjrkman2012} &   $-$0.020$\pm$0.001\cite{Zacharia2004}  & 23.0\cite{Bjrkman2012}  & 37$-$41\cite{Landolt2003,Rydberg2003} \\
                 & $-$0.27\cite{Bjrkman2012} & $-$0.016$\pm$0.001\cite{Liu2011,Liu2012}   & 46.1\cite{Bjrkman2012} &\\
                 & $-$0.29\cite{Lebegue2010} & $-$0.013$\pm$0.005\cite{Benedict1998}  & 13\cite{Rydberg2003} &\\
				  & $-$0.34\cite{Spanu2009} &   &  & \\
				  & $-$0.34$^*$ &   & 15.19$^*$ & \\
				  \hline
h-BN        & $-$0.36 \cite{Bjrkman2012} &    & 20.2\cite{Bjrkman2012} &\\
				 & $-$0.26 \cite{Bjrkman2012} &   & 41.2\cite{Bjrkman2012} &\\
				 &                                             &   & 11\cite{Rydberg2003} &\\ 
                & $-$0.52$^*$ &  & 12.10$^*$ &\\
                
\hline            
MoS$_2$        & $-$0.44\cite{Bjrkman2012}   &  &  24\cite{Bjrkman2012}  &\\
                       & 											  &  &  55.2\cite{Bjrkman2012} &\\
                      & 												 &  &  49\cite{Rydberg2003} &\\
                       & $-$0.37$^*$ &  & 31.27$^*$  &\\

\hline
\hline

\begin{tablenotes}
      \item *This work.
    \end{tablenotes}

\end{tabular}
\end{table*}

\begin{table}[h]
\caption{\label{R2IE} R$^{2}$, Root Mean Squared Error (MRSE)  and mean absolute error (MAE) on test, train, and validations set for interlayer energy and C$_{33}$. Values of MSE and MAE are in eV/{\AA}$^2$ for IE,  and in GPa for C$_{33}$. }
\begin{ruledtabular}   
\begin{tabular}{lccccc}
Set      &  R$^2$ &  RMSE  & MAE \\
\hline
IE-BNN-Test-Set                           &0.80   &  0.055  &0.035 \\
IE-BNN-Train-Set                           &0.97   &    0.014   &0.010  \\
IE-BNN-Valid-Set                        &0.72   &    0.089   &0.055  \\
\hline
C$_{33}$-BNN-Test-Set                           &0.80    & 9.98  &16.04  \\
C$_{33}$-BNN-Train-Set                           &0.98     & 5.99  &5.76  \\
C$_{33}$-BNN-Valid-Set                          &0.73     & 11.89  &20.65  \\

\end{tabular}
\end{ruledtabular}
\end{table}

\begin{figure*}[ht]
\centering
  \begin{tabular}{@{}cccc@{}}
     \includegraphics[width=.7\textwidth]{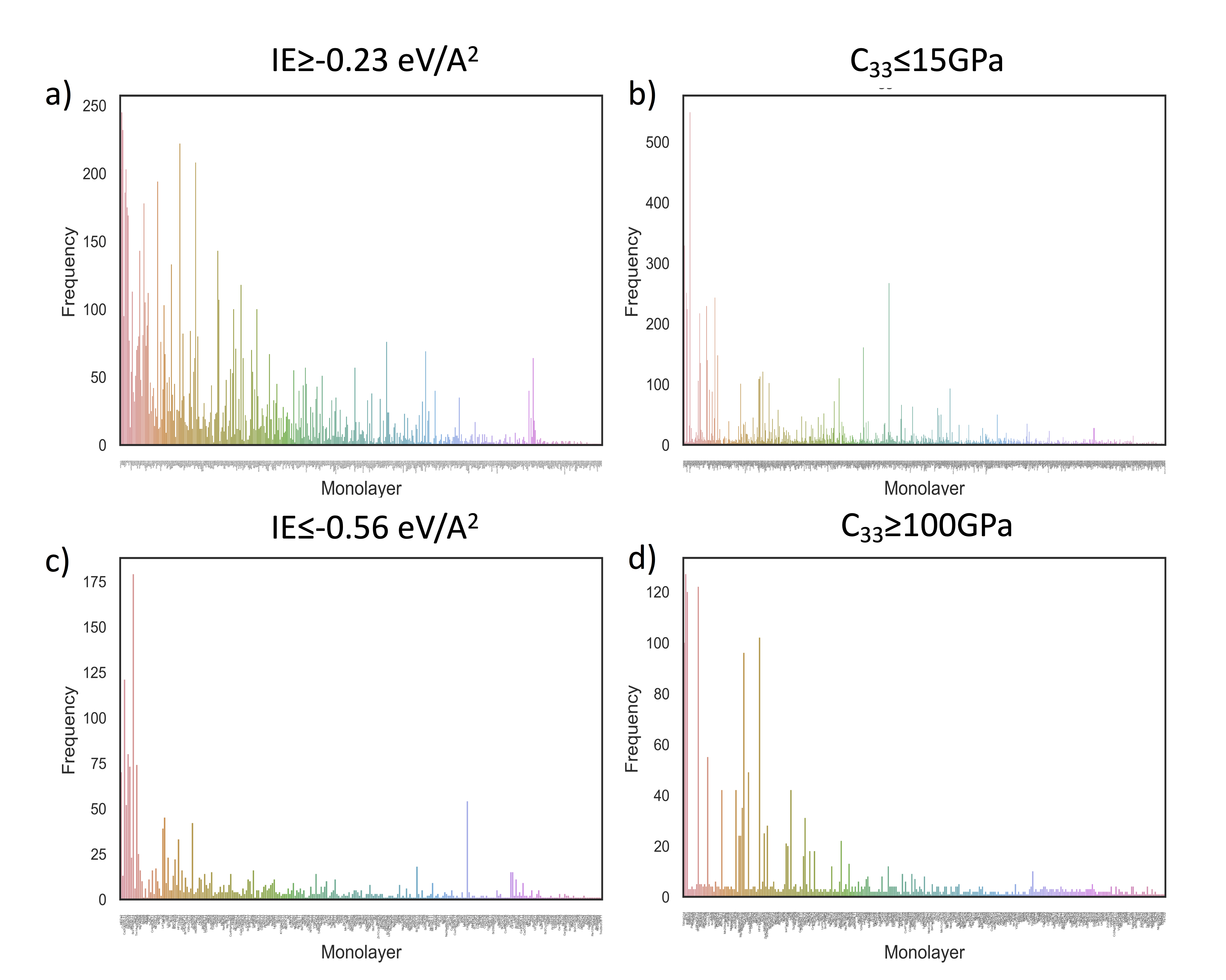} \\
  \end{tabular}
  \caption{(Color online) Frequency of the monolayers that appear in the first 5000 bilayers with a) IE $\geq-$0.23 eV/{\AA} or b) C$_{33}$ $\leq15.00$ GPa, and the ones that appear in the first 1000 bilayers with c) IE $\leq-$0.56 eV/{\AA} or d) C$_{33}$ $\geq100$ GPa.}
  \label{fig:pic2}
\end{figure*}

\begin{table}[h]
\caption{\label{FreqL} Monolayer frequency among the bilayers with small and large IE and C$_{33}$ values, included in the $\sim$300k set of stable structures. In this analysis, we consider the first 5000 biayers with IE $\geq-$0.23 eV/{\AA} or C$_{33}$ $\leq$15.00 GPa, and the first 1000 bilayers with IE   $\leq-$0.56 eV/{\AA} or   C$_{33}$  $\geq$100.00 GPa. 
The ``T1'' prefix denotes the T1 polymorph of transition metal chalcogenides. Here, we use the stoichiometric notation of the individual monolayers as in the https://materialsweb.org database.\cite{Materialsweb}}

\begin{ruledtabular}   
\begin{tabular}{lccccc}

&   IE $\geq-$0.23 eV/{\AA} &  C$_{33}$ $\leq$15.00 GPa\\

\hline
            & In$_2$S$_2$: 245 &          ZrCd$_2$H$_{12}$O$_{6}$F$_{8}$: 550\\
            &  As$_4$S$_6$: 233 &        CdO: 329 \\ 
             &  In$_2$Cl$_2$O$_2$: 222 &           Cd$_2$P$_2$S$_6$: 267\\ 
            &  Ga$_2$Se$_2$: 208 &            BN-T1: 251\\ 
& In$_2$Se$_2$: 204  &                        BN: 243\\
& In$_2$Se$_2$-T1: 203  &                C$_2$: 229\\
& Tc$_4$P$_{16}$: 194  &                  Cd$_2$Te$_2$Mo$_2$O$_{12}$: 224\\
& Sc$_2$P$_2$S$_8$: 186  &                Cd$_2$P$_2$S$_6$-T1: 217\\
& Al$_2$Cl$_6$: 178  &                       Al$_4$Cd$_2$Cl$_{16}$: 161\\
\hline
\hline
&   IE $\leq-$0.56 J/m$^2$ &  C$_{33}$ $\geq$100.00 GPa\\
\hline
            & Sr$_2$Ti$_2$Si$_4$O$_{14}$:179 &          HfFeCl$_6$-T1: 89\\
            &  Ca$_2$La$_2$I$_{10}$-T1: 121 &        BN: 83 \\ 
             &  Ti$_2$Ge$_2$O$_6$: 81 &           Sr$_2$Ti$_2$Si$_4$O$_{14}$: 75\\ 
            &  TmAgP$_2$Se$_6$: 74 &            BN-T1: 44\\ 
& Ti$_6$H$_4$O$_{14}$: 73 &                        Si$_4$O$_8$: 40\\
& Hf$_3$Te$_2$: 70  &                Al$_2$Si$_2$H$_4$O$_9$: 35\\
& Ca$_2$La$_2$I$_{10}$: 54  &                  Ti$_6$H$_4$O$_{14}$: 30\\
& Ta$_3$TeI$_7$: 52  &                Si$_4$O$_8$-T1: 28\\
& V$_4$F$_{16}$: 45 &                       Al$_2$Si$_4$O$_{11}$: 26\\
\hline

\end{tabular}
\end{ruledtabular}
\end{table}

Although the 6,138 monolayers were theoretically predicted from the existing and thermodynamically relative bulk counterpart, the decomposition energy suggests that most of these structures are not stable, which implies that the stability of a large fraction of the 18,834,453 bilayers cannot be ensured at standard conditions.\cite{Zhou2019}  Indeed, only   3,497 monolayers have a decomposition energy larger than 100 meV, the typical threshold for metastability, which can generate 6,114,504 bilayers. However, is has been shown that  a large number of structures are stable although its decomposition energy was measured to be considerably lower than the threshold. \cite{Sun2016}

We continue our analysis on a subset of the $\sim$18M bilayers that have a decomposition energy $\geq$0 meV, and an exfoliation energy $\leq$55 meV,  which is estimated to be lower than existing 2D materials, suggesting the possibility of exfoliation  from bulk phases.\cite{WangNC2015,Ashton2016,Ashton2017,Materialsweb}  This new subset consists of 770 monolayers and show a distance to their respective thermodynamic convex hulls that indicates thermodynamically stability  at standard conditions.\cite{Ashton2016,Ashton2017}  Here, The 3-dimensional counterpart structure of the 770 monolayers were selected among a large number of thermodynamically stable structures formed by the elements listed in the Section Methods. 
A subset of bulk materials possessing layered geometries in their crystal lattice were identified by using a topology-scaling algorithm, which measures the sizes of bonded atomic clusters in a structure's unit cell, and determines their scaling with cell size. It has been estimated that a larger part (around 50$\%$) of 2D counterpart forms with the stoichiometry ABC, AB$_2$, AB, AB$_3$, and ABC$_2$.\cite{Ashton2016,Ashton2017} The combinations of the 770 monolayers generate 296,835 of stable and manufacturable bilayers.  
Using our BNN model, we extrapolated the interlayer energy and C$_{33}$ for the thermodynamically stable $\sim$300k bilayers, together with the temperature stability, shown in Figures \ref{fig:pic1} (the complete list of values is reported available online (http://doi.org/10.26195/5dd36650d7e1e). The quality of the results is assured by the fact that the materials in the very large virtual screening set lie within or close to the domain of applicability of the two models. 
Interestingly, although the IE and C$_{33}$ values follow logarithmic distribution in the $\sim$18M set and a linear distribution in the $\sim$300k subset, the calculated Pearson and  Spearman for the two sets are very close, confirming the lack of correlation between the two properties. 
From a screening of the dataset $\sim$300k structures, we extracted the most common monolayers with a low absolute value of the interlayer  energy ($\geq-$0.23 eV/{\AA}$^2$) and low C$_{33}$ ($\leq$15.00 GPa), represented in Figure \ref{fig:pic2}, whereas  the top nine most frequent monolayers are listed in Table \ref{FreqL}. Figure \ref{fig:pic1} shows areas, at around 550 on Monolayer2 axis, where a few monolayers, when coupled with a large fraction of the whole set of monolayers, will form a bilayer with a particularly high C$_{33}$. We found the most frequent monolayer to give high C$_{33}$ to be HfFeCl$_6$-T1 (see Table \ref{FreqL}), which is the sixth most frequent monolayer with high C$_{33}$  in the $\sim$18M set. Interestingly, BN appear frequently in both high and low C$_{33}$ bilayers. In conclusion, our IE predictions suggest the possible application of a significant fraction of the $\sim$300k bilayers as super-lubricant, where the best one would be as a As$_4$S$_6$-In$_2$Se$_2$ together with its polymorph form As$_4$S$_6$-In$_2$Se$_2$-T1 with and interlayer energy of $-$0.12 eV/{\AA}$^2$, which however is stable only up to 69K, while its polymorph form As$_4$S$_6$-In$_2$Se$_2$-T1, while having a very similar interlayer energy,  is stable up to 928K. Furthermore the As$_4$S$_6$-In$_2$Se$_2$ predicted C$_{33}$ value of the As$_4$S$_6$-In$_2$Se$_2$  is 39.49 GPa, which is relatively low, ensuring its wide applicability. On the other hand  ZrCd$_2$H$_{12}$O$_{6}$F$_{8}$-Hf$_2$Br$_2$N$_2$-T1 is the softest bilayer, with a C$_{33}$ value of 4.04 GPa. Although the  ZrCd$_2$H$_{12}$O$_{6}$F$_{8}$-Hf$_2$Br$_2$N$_2$-T1 has interlayer energy value of $-$0.36 eV/{\AA}$^2$, which it lies in the middle of the our IE range, it represents the most universal dry lubricant. 
Thoroughly, we calculated to be 287,245 structure with decomposition energy $>$0 meV/atom, 60,910 of which are stable at room temperature, and 19,900 up to 1300K.



With the present work, we created a very large database of atomic properties (IE and C$_{33}$, together with the relative temperature stability) for a class of materials with growing technological and scientific interest.  Furthermore, we demonstrated the potential of machine learning in amplifying the  capabilities of conventional computational approaches used in materials discovery. The approach described, which here is use coupled by DFT calculations, is fully transferable to other the scenarios where an elevated number of structures can be generated by the combinations of a relatively few atomic structures.

\section{Methodology}

Our Machine Learning approach relies on amounts of high quality structured data, using a set  of descriptors to indicate known properties, from which the algorithm will learn hidden patterns. In general, the problem can be reduced to the identification of a general non-linear function $Y=f(X)$, where here $Y$ is represent the interlayer energy and the elastic constant (C$_{33}$), $X$ represent the input space of descriptors, and $f$ is the transfer function that link the descriptors to the response variable.  
The work-flow of our implementation is structured in three main parts, concatenated as follows:

	A) Data Collection A.1) Density functional theory calculations
	B) Data Preparation  B.1) Feature selection B.2) Cluster analysis (for input data randomization)
	C) Bayesian Neural Network (model optimization) 
	D) Statistical Analysis.

\subsection{Data Collection}

\subsubsection{Density functional theory calculations}

\begin{figure}[ht]
\centering
  \begin{tabular}{@{}cccc@{}}
     \includegraphics[width=.35\textwidth]{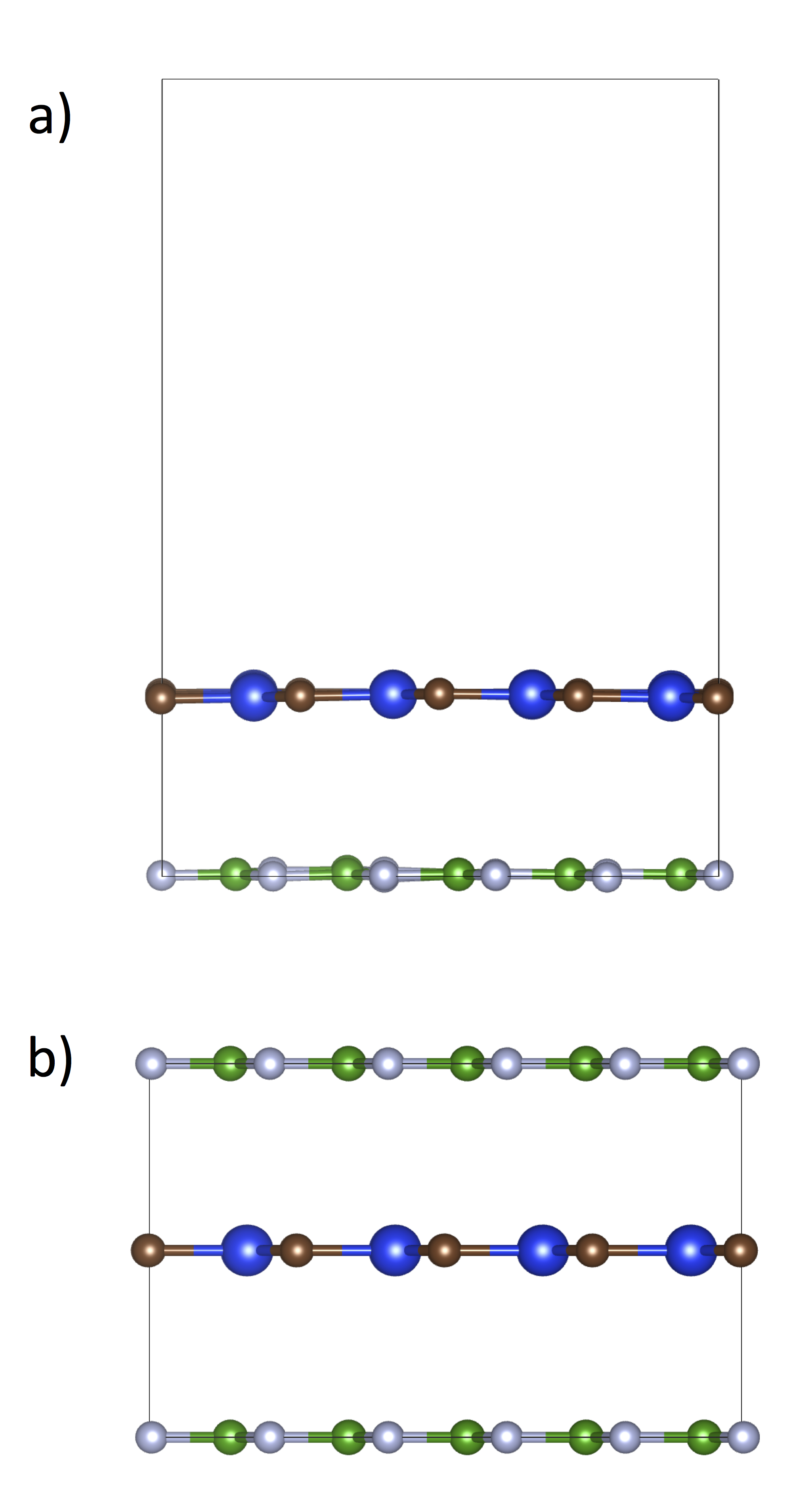} \\
  \end{tabular}
  \caption{ (Color online). a) and b) figures schematically represent the supercell used in the DFT calculations of the interlayer energy and C$_{33}$, respectively (i.e. BN$-$SiC). For the calculation of the interlayer energy, the size of the supercell along the z-axis is large enough to prevent interactions between replica of the layers in periodic boundary conditions, whereas for the calculation of C$_{33}$ the dimension along the z-axis is the equilibrium distance of the two monolayers.}
  \label{fig:supercell}
\end{figure}

To calculate the interlayer energy of the 282 and the C$_{33}$ of the 226 bilayer structures by means of DFT, we used VASP within the GGA-PBE approximation where a Tkatchenko-Scheffler van der Waals correlation correction was applied.\cite{KresseJPCM1994,KressePRB1999,Tkatchenko2009} A k-point space of 8$\times$8$\times$1 for structures with atoms less than 10, and 3$\times$3$\times$1 otherwise, and an energy cut-off is 520 eV. The energy minimization tolerance is 10$^{-6}$ eV, and the force tolerance is 10$^{-2}$ eV/{\AA}. 
We calculated the interlayer energy as the difference between the total energy of the individual monolayers and the total energy of the bilayer, where a negative energy indicates attractive interaction, then normalize this quantity per unit area. The supercell size along the z-axis was chosen to be large enough to avoid interactions with replica of the layer in the periodic boundary conditions. To calculate the value of the elastic constant, we consider its value along the z-axis of a bilayer (i.e. C$_{33}$). Due to the different forces acting in-plane and out-of-plane, we can approximate the hardness with the C$_{33}$.\cite{Bertolazzi2011,Liu2016} We calculate the C$_{33}$ by interpolating the interlayer energy change as a function of the interlayer distance, obtained by varying the supercell size along the z-axis. Here, the the supercell dimension at the equilibrium along the z-axis doubles the equilibrium distance of the bilayers previously calculated, to resemble a two-component multi-layered structure, as schematically shown in Figure \ref{fig:supercell}. When varying the supercell size, we remain in the elastic domain of the solid and a quadratic dependence of the total energy with respect to the strain is expected.

\subsubsection{Descriptors}

We obtained structural information for 6,138 monolayers from an online database (https://2dmatpedia.org/). The 6,138 monolayers were obtained from a large number of inorganic bulk structures available on the online database Materials Project (https://materialsproject.org/) by using a ``top-down'' approach where the bulk crystals are screened for layered structures which are then theoretically exfoliated to 2D monolayers, and a ``bottom-up'' approach, in which elemental substitution is systematically applied to the unitary and binary 2D materials obtained from the top-down approach. 
To obtain feature vector for materials  ($X$) we adapted a method we developed previously, which allows us to calculate materials fragment descriptors computed from the connectivity graph inside the unit cell.\cite{Isayev2017}
Within this approach, each crystal structure is represented as a graph, with vertices decorated according to the reference properties of the atoms they represent, and each node is connected to its neighbour according to the Voronoi tessellation. The adjacency matrix of this graph determines the global topology for a given system, including interatomic bonds and contacts within a crystal. The final descriptor vector for the machine learning  model is obtained by partitioning a full graph into subgraphs called fragments. 
Descriptors for each bilayer were obtained by adding the values of the descriptors for the two monolayers, as described by Tawfik et al.\cite{Tawfik2019}

\subsection{Data Preparation}

\subsubsection{Features Selection}

The number of descriptors calculated for our structures is too large to be used for our calculations, as overfitting will occur. Feature selection involves choosing a subset of $d$ features from a set of $D$ features based using some optimization criterion, creating a more compact descriptor space $X$ with as little performance loss as possible. The features removed should therefore be largely irrelevant for the calculation of a specific target property. For this purpose, we use a combination of Genetic Algorithm (GA) search and LASSO regression.\cite{Tibshirani1996} The idea of GA is to generate some random possible solutions, which represent different variables, to then combine the best solutions in an iterative process. The GA process tries to maximise a fitness function, that in our case is the LASSO function. We further screen the features using a LASSO regression analysis. The goal of LASSO regression is to obtain the subset of descriptors ($X$) that minimizes prediction error for a quantitative response variable ($Y$). The LASSO does this by imposing a constraint on the model parameters that causes regression coefficients for some variables to drop to zero. Variables with a regression coefficient equal to zero after the shrinkage process are excluded from the model. Variables with non-zero regression coefficients variables are most strongly associated with the response variable. The goal of the algorithm is to minimize:

\begin{equation}
\label{LASSO}
L=\sum^{n}_{i=1} (y_i-\sum x_{ij}\beta_{j})^2 + \lambda \sum^{p}_{j=1}|\beta_{j}|
\end{equation}

Where the tuning parameter, $\lambda$ controls the strength of the penalty.  Therefore, $\lambda$ control the degree of elimination:
When $\lambda = 0$, no parameters are eliminated. The estimate is equal to the one found with linear regression.
As $\lambda$ increases, more and more coefficients are set to zero and eliminated (theoretically, when  $\lambda = \infty$, all coefficients are eliminated).
As $\lambda$ increases, bias increases.
As $\lambda$ decreases, variance increases.

\subsubsection{Cluster Analysis}

The choice of representative structures to be used to train our model is crucial for the quality of predictions. K-means is a method of vector quantization that is popular for cluster analysis in data mining.\cite{Forgy1965} K-means clustering aims to partition n observations into k clusters in which each observation belongs to the cluster with the nearest mean, acting as a representative model of the cluster. This results in a partitioning of the data space into Voronoi cells.
The best number of clusters k leading to the largest distance is not known a priori and must be computed from the data. The objective of K-means clustering is to minimize total intra-cluster variance, or, the squared error function:

\begin{equation}
\label{KMEANS}
J=\sum^{k}_{j=1}\sum^{n}_{i=1}||x^{i}_{j}-c_{j}||^2 \quad .
\end{equation}

This procedure will maximize the diversity of structures assigned to training and test sets while ensuring that the test sets are still within the domain of the models. Here, we use the silhouette score to express the quality of the clustering. The silhouette score, with values between +1 and $-$1, is a measure to indicate how close each point in one cluster is to points in the neighbouring clusters, where +1 indicates that the sample is far away from the neighbouring clusters. A K-means analysis gives the best score when three sub-groups are formed for interlayer energy, with an average silhouette score of 0.78, and five for C$_{33}$ silhouette score of 0.63. The training set contains 75$\%$ of the data and the test set 25$\%$.

\subsection{Bayesian Neural Networks} \label{BNN}
In the present work we use machine learning in a Bayesian framework in order to predict not only the transfer function and the property of a large number of structures, but also to give the confidence interval for each value.\cite{Bergmann2014}
In the Bayesian point of view, regressions are formulated using probability distributions rather than point estimates. The target property or response, $Y$, is not estimated as a single value, but is assumed to be drawn from a probability distribution. 
The aim of Bayesian regressions is not to find the single ``best'' value of the model parameters, but rather to determine the posterior distribution for the model parameters.\cite{MacKay1995,MacKay1992,Burden1999} Not only is the response generated from a probability distribution, but the model parameters are assumed to come from a distribution as well. The posterior probability of the model parameters is conditional upon the train inputs and outputs:

\begin{eqnarray}
P(\beta | y, X) = \frac{P(y| \beta, X) \times P(\beta | X)}{P(y|X)}  \quad ,
\end{eqnarray}

Here, P($\beta$ $|$ y, X) is the posterior probability distribution of the model parameters given the inputs and outputs. This is equal to the likelihood of the data, P(y $|$ $\beta$, X), multiplied by the prior probability of the parameters and divided by a normalization constant. 
%
%

Here, we have a posterior distribution for the model parameters that is proportional to the likelihood of the data multiplied by the prior probability of the parameters. We can observe two primary benefits of Bayesian regressions. Priors: parameters distributions are included in the model. If these are unknown, we can use non-informative priors for the parameters such as a normal distribution. Posterior: the result of performing Bayesian regression is a distribution of possible model parameters based on the data and the prior. This allows us to quantify our uncertainty about the model.

To implement this methodology, we use a dropout approach, which can be seen as a Bayesian approximation of a well known probabilistic model. Dropout is used in many models in deep learning as a way to avoid over-fitting by randomly creating deviations from the optimizaton pathway. In our implementation dropout approximately integrates over the weights in the model.\cite{Gal2016,Srivastava2014,Tran2019} Basically, this is comparable to performing a number of stochastic passes through the network, and then averaging the results. This result has been presented in the literature before as model averaging. With dropout, we sample binary variables for every input point and for every network unit in each layer. Each binary variable takes value 1 with probability p$_i$ for layer i. A unit is dropped to zero for a given input if its corresponding binary variable takes value 0. We use the same values in the backward pass propagating the derivatives to the parameters, obtaining a distribution over each descriptor, from which we can extrapolate the uncertainty associated to each prediction.

We use a BNN with 2 hidden layers composed of 128 neurons each, where the dropout probability is 0.1. The dropout creates a distribution over the calculated response, which is then averaged over 600 trial networks giving the response value and the associate standard deviation. 

\begin{figure*}[ht]
\centering
  \begin{tabular}{@{}cccc@{}}
     \includegraphics[width=.54\textwidth]{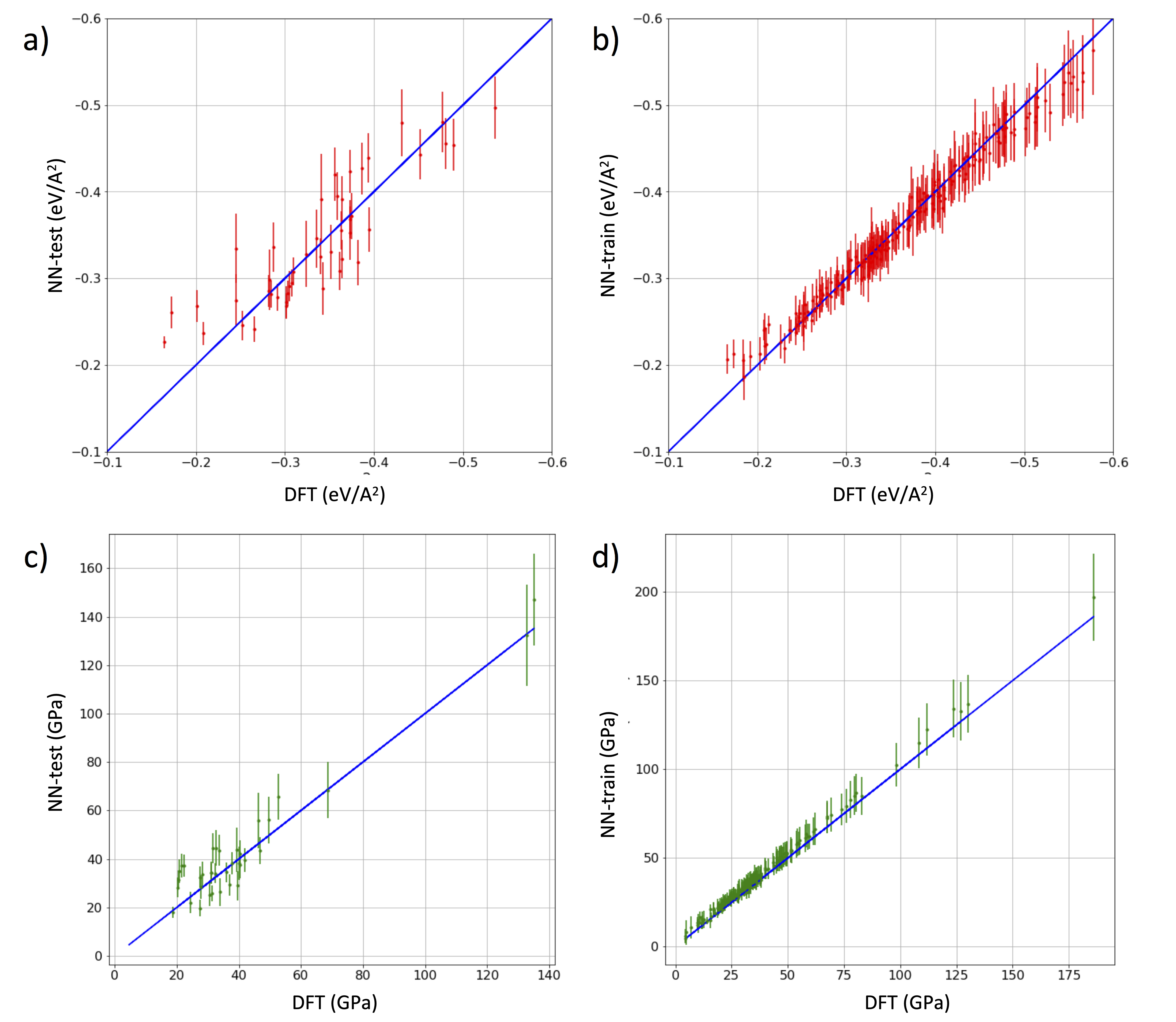} \\
  \end{tabular}
  \caption{(Color online). a) and b) Interlayer energy and associated Bayesian uncertainty of the test and train set, respectively, used to create our ML model. c) and d) C$_{33}$ and associated Bayesian uncertainty of the test and train, respectively, set used to create our ML model.}
  \label{fig:BNNTRAINTEST}
\end{figure*}

Figure \ref{fig:BNNTRAINTEST} shows the IE and C$_{33}$ values, with relative error-bar, of the train and test set after BNN optimization.

\subsection{Statistical Analysis}

To investigate correlations between the properties, we calculated the Pearson product moment correlation, which evaluates the linear relationship between two continuous variables, and Spearman rank-order correlation, which evaluates the monotonic relationship between two continuous or ordinal variables, as follows:

\begin{equation}
\label{pearson}
r = \frac{{}\sum_{i=1}^{n} (x_i - \overline{x})(y_i - \overline{y})}
{\sqrt{\sum_{i=1}^{n} (x_i - \overline{x})^2(y_i - \overline{y})^2}}  \quad ,
\end{equation}

where $x_i$ and $y_i$ are the two variables, and

\begin{equation}
\label{Spearman}
\rho = 1- {\frac {6 \sum d_i^2}{n(n^2 - 1)}} \quad ,
\end{equation}

where $d =$ is  the pairwise distances of the ranks of the variables $x_i$ and $y_i$, and $n =$ is  the number of samples.
Both coefficients are very small in both DFT and ML results, indicating lack of correlation between IE and C$_{33}$.
Although there is no correlation between the two quantities, we calculated that 90$\%$ of the bilayers have an IE between $-$0.51 eV and $-$0.28 eV and a C$_{33}$ between 19.44 GPa and 63.44 GPa, as reported in Table \ref{PERC_IE_C33}.

\begin{table}[h]
\caption{\label{PERC_IE_C33} Distribution of the IE and C$_{33}$ across the $\sim$300k and $\sim$18M datasets. On each line is indicated the percentage of bilayers with IE or C$_{33}$ values larger than the one shown in the corresponding column.  }
\begin{ruledtabular}   
\begin{tabular}{lccccc}
      &  $\sim$300k &    & $\sim$18M & \\
Percentage      &  IE (eV) &  C$_{33}$ (GPa) & IE (eV) & C$_{33}$ (GPa)\\
\hline
0.05$\%$                           &$-$0.48   & $-$18.80  &$-$0.51 &19.44  \\
0.50$\%$                          &$-$0.35   & $-$34.96  &$-$0.38 &34.86  \\
0.95$\%$                           &$-$0.25   & $-$69.07  &$-$0.28 &63.44  \\

\end{tabular}
\end{ruledtabular}
\end{table}

\section{List of Elements} \label{elements}
List of elements that form stable bilayers:
Ag,
Al,
As,
Au,
B,
Ba
Bi,
Br,
C,
Ca,
Cd,
Ce,
Cl,
Co,
Cr,
Cu,
Dy,
Er,
Eu,
F,
Fe,
Ga,
Gd,
Ge,
H,
Hf,
Hg,
Ho,
I,
In,
Ir,
K,
La,
Li,
Lu,
Mg,
Mn,
Mo,
N,
Na,
Nb,
Nd,
Ni,
Np,
O,
Os,
P,
Pa,
Pb,
Pd,
Pr,
Pt,
Pu,
Rb,
Re,
Rh,
Ru,
S,
Sb,
Sc,
Se,
Si,
Sm,
Sn,
Sr,
Ta,
Tb,
Tc,
Te,
Th,
Ti,
Tl,
Tm,
U,
V,
W,
Y,
Yb,
Zn,
Zr.

\section{Data availability}

The complete IE, C$_{33}$ and thermal stability data is available online (http://doi.org/10.26195/5dd36650d7e1e).

\section{Code availability}

Custom Python codes for data preprocessing and Bayesian Neural Network training and data extrapolation are available online (https://github.com/fronzi/projBNN). 


\section{Competing interests}

The authors declare no competing interests.

\section{Acknowledgements}

The authors gratefully acknowledge the financial support of the National Natural Science Foundation of China (No. 51323011), and the Australian Government through the Australian Research Council (ARC DP16010130). The theoretical calculations in this research were undertaken with the assistance of resources from the National Computational Infrastructure (NCI), which is supported by the Australian Government. The theoretical calculations in this work were also supported by resources provided by the Pawsey Supercomputing Centre with funding from the Australian Government and the Government of Western Australia.



\begin{thebibliography}{10}
\expandafter\ifx\csname url\endcsname\relax
  \def\url#1{\texttt{#1}}\fi
\expandafter\ifx\csname urlprefix\endcsname\relax\def\urlprefix{URL }\fi
\providecommand{\bibinfo}[2]{#2}
\providecommand{\eprint}[2][]{\url{#2}}

\bibitem{Sliney1978}
\bibinfo{author}{Sliney, H.~E.}
\newblock \bibinfo{title}{Dynamics of solid lubrication as observed by optical
  microscopy}.
\newblock \emph{\bibinfo{journal}{ASLE Trans.}} \textbf{\bibinfo{volume}{21}},
  \bibinfo{pages}{109--117} (\bibinfo{year}{1978}).

\bibitem{Filippov2013}
\bibinfo{author}{Filippov, A.~E.}, \bibinfo{author}{Dienwiebel, M.},
  \bibinfo{author}{Frenken, J. W.~M.}, \bibinfo{author}{Kla, J.} \&
  \bibinfo{author}{Urbakh, M.}
\newblock \bibinfo{title}{Torque and twist against superlubricity}.
\newblock \emph{\bibinfo{journal}{Phys. Rev. Lett.}}
  \textbf{\bibinfo{volume}{100}}, \bibinfo{pages}{046102}
  (\bibinfo{year}{2008}).

\bibitem{Novoselov2016}
\bibinfo{author}{Novoselov, K.~S.}, \bibinfo{author}{Mishchenko, A.},
  \bibinfo{author}{Carvalho, A.} \& \bibinfo{author}{Neto, A. H.~C.}
\newblock \bibinfo{title}{2d materials and van der Waals heterostructures}.
\newblock \emph{\bibinfo{journal}{Science}} \textbf{\bibinfo{volume}{353}},
  \bibinfo{pages}{9439} (\bibinfo{year}{2016}).

\bibitem{Geim2013}
\bibinfo{author}{Geim, A.~K.} \& \bibinfo{author}{Grigorieva, I.~V.}
\newblock \bibinfo{title}{Van der Waals heterostructures}.
\newblock \emph{\bibinfo{journal}{Nature}} \textbf{\bibinfo{volume}{499}},
  \bibinfo{pages}{419} (\bibinfo{year}{2013}).

\bibitem{Wang2018}
\bibinfo{author}{Wang, J.} \emph{et~al.}
\newblock \bibinfo{title}{Ultralow interlayer friction of layered electride
  Ca$_2$N: A potential two-dimensional solid lubricant material}.
\newblock \emph{\bibinfo{journal}{Materials}} \textbf{\bibinfo{volume}{11}},
  \bibinfo{pages}{2462} (\bibinfo{year}{2018}).

\bibitem{Jain2013}
\bibinfo{author}{Jain, A.} \emph{et~al.}
\newblock \bibinfo{title}{The materials project: A materials genome approach to
  accelerating materials innovation}.
\newblock \emph{\bibinfo{journal}{APL Materials}} \textbf{\bibinfo{volume}{1}},
  \bibinfo{pages}{011002} (\bibinfo{year}{2013}).

\bibitem{Zhou2019}
\bibinfo{author}{Zhou, J.} \emph{et~al.}
\newblock \bibinfo{title}{2dmatpedia, an open computational database of
  two-dimensional materials from top-down and bottom-up approaches}.
\newblock \emph{\bibinfo{journal}{Scientific Data}} \textbf{\bibinfo{volume}{6}}, \bibinfo{pages}{86}
  (\bibinfo{year}{2019}).

\bibitem{Ponomarenko2011}
\bibinfo{author}{Ponomarenko, L.~A.} \emph{et~al.}
\newblock \bibinfo{title}{Tunable metal-insulator transition in double-layer
  graphene heterostructures}.
\newblock \emph{\bibinfo{journal}{Nature Phys.}} \textbf{\bibinfo{volume}{7}},
  \bibinfo{pages}{958--961} (\bibinfo{year}{2011}).

\bibitem{Britnell2012}
\bibinfo{author}{Britnell, L.} \emph{et~al.}
\newblock \bibinfo{title}{Field-effect tunneling transistor based on vertical
  graphene heterostructures}.
\newblock \emph{\bibinfo{journal}{Science}} \textbf{\bibinfo{volume}{335}},
  \bibinfo{pages}{947--950} (\bibinfo{year}{2012}).

\bibitem{Haigh2012}
\bibinfo{author}{Haigh, S.~J.} \emph{et~al.}
\newblock \bibinfo{title}{Cross-sectional imaging of individual layers and
  buried interfaces of graphene-based heterostructures and superlattices}.
\newblock \emph{\bibinfo{journal}{Nature Mater.}}
  \textbf{\bibinfo{volume}{11}}, \bibinfo{pages}{764--767}
  (\bibinfo{year}{2012}).

\bibitem{Georgiou2013}
\bibinfo{author}{Georgiou, T.} \emph{et~al.}
\newblock \bibinfo{title}{Vertical field-effect transistor based on
  graphene-WS$_2$ heterostructures for flexible and transparent electronics}.
\newblock \emph{\bibinfo{journal}{Nature Nanotechnol.}}
  \textbf{\bibinfo{volume}{8}} (\bibinfo{year}{2013}).

\bibitem{Wang2013}
\bibinfo{author}{Wang, L.} \emph{et~al.}
\newblock \bibinfo{title}{One-dimensional electrical contact to a
  two-dimensional material}.
\newblock \emph{\bibinfo{journal}{Science}} \textbf{\bibinfo{volume}{342}},
  \bibinfo{pages}{614--617} (\bibinfo{year}{2013}).
  
  
  \bibitem{Goedecker1999}
\bibinfo{author}{Goedecker, S.}
\newblock \bibinfo{title}{Linear scaling electronic structure methods}.
\newblock \emph{\bibinfo{journal}{Rev. Mod. Phys.}}
  \textbf{\bibinfo{volume}{71}}, \bibinfo{pages}{1085} (\bibinfo{year}{1999}).

\bibitem{Bowler2012}
\bibinfo{author}{Bowler, D.} \& \bibinfo{author}{Miyazaki, T.}
\newblock \bibinfo{title}{O(n) methods in electronic structure calculations}.
\newblock \emph{\bibinfo{journal}{Rep. Prog. Phys.}}
  \textbf{\bibinfo{volume}{75}}, \bibinfo{pages}{036503}
  (\bibinfo{year}{2012}).

\bibitem{Correa2009}
\bibinfo{author}{Correa, M.}, \bibinfo{author}{Bielza, C.} \&
  \bibinfo{author}{Pamies-Teixeirac, J.}
\newblock \bibinfo{title}{Comparison of bayesian networks and artificial neural
  networks for quality detection in a machining process}.
\newblock \emph{\bibinfo{journal}{Expert Systems with Applications}}
  \textbf{\bibinfo{volume}{36}}, \bibinfo{pages}{7270--7279}
  (\bibinfo{year}{2009}).

\bibitem{Butler2018}
\bibinfo{author}{Butler, K.~T.}, \bibinfo{author}{Davies, D.~W.},
  \bibinfo{author}{Cartwright, H.}, \bibinfo{author}{Isayev, O.} \&
  \bibinfo{author}{Walsh, A.}
\newblock \bibinfo{title}{Machine learning for molecular and materials
  science}.
\newblock \emph{\bibinfo{journal}{Nature}} \textbf{\bibinfo{volume}{559}},
  \bibinfo{pages}{547--555} (\bibinfo{year}{2018}).

\bibitem{Lu2017}
\bibinfo{author}{Lu, N.} \emph{et~al.}
\newblock \bibinfo{title}{Twisted MX$_2$/MoS$_2$ heterobilayers: effect of van der
  Waals interaction on the electronic structure}.
\newblock \emph{\bibinfo{journal}{Nanoscale}} \textbf{\bibinfo{volume}{9}},
  \bibinfo{pages}{19131} (\bibinfo{year}{2017}).



\bibitem{Bertolazzi2011}
\bibinfo{author}{Bertolazzi, S.}, \bibinfo{author}{Brivio, J.} \&
  \bibinfo{author}{Kis, A.}
  \newblock \bibinfo{title}{Stretching and Breaking of Ultrathin MoS$_2$}.
\newblock \emph{\bibinfo{journal}{ACS nano}} \textbf{\bibinfo{volume}{5}},
  \bibinfo{pages}{9703} (\bibinfo{year}{2011}).


\bibitem{Liu2016}
\bibinfo{author}{Liu, K.} \& \bibinfo{author}{Wu, J.}
\newblock \bibinfo{title}{Mechanical properties of two-dimensional materials and heterostructures}.
\newblock \emph{\bibinfo{journal}{Journal of Materials Research}}
  \textbf{\bibinfo{volume}{31}}, \bibinfo{pages}{832} (\bibinfo{year}{2016}).

\bibitem{Bjrkman2012}
\bibinfo{author}{Bj\"{o}rkman, T.}, \bibinfo{author}{Gulans, A.},
  \bibinfo{author}{Krasheninnikov, A.~V.} \& \bibinfo{author}{Nieminen, R.~M.}
\newblock \bibinfo{title}{van der Waals bonding in layered compounds from
  advanced density-functional first-principles calculations}.
\newblock \emph{\bibinfo{journal}{Phys. Rev. Lett.}}
  \textbf{\bibinfo{volume}{108}}, \bibinfo{pages}{235502}
  (\bibinfo{year}{2012}).

\bibitem{Rego2015}
\bibinfo{author}{R\^{e}go, C. R.~C.}, \bibinfo{author}{Oliveira, L.~N.} \&
   \bibinfo{author}{Tereshchuk, ~P.},\bibinfo{author}{{Da Silva}, J. L. F.} 
\newblock \bibinfo{title}{Comparative study of van der Waals corrections to the
  bulk properties of graphite}.
\newblock \emph{\bibinfo{journal}{J. Phys.: Condens. Matter}}
  \textbf{\bibinfo{volume}{27}}, \bibinfo{pages}{415502}
  (\bibinfo{year}{2015}).

\bibitem{Bjorkman2012b}
\bibinfo{author}{Bj\"{o}rkman, T.}, \bibinfo{author}{Gulans, A.},
  \bibinfo{author}{Krasheninnikov, A.~V.} \& \bibinfo{author}{Nieminen, R.~M.}
\newblock \bibinfo{title}{Are we van der Waals ready?}
\newblock \emph{\bibinfo{journal}{Journal of Physics: Condensed Matter}}
  \textbf{\bibinfo{volume}{24}}, \bibinfo{pages}{424218}
  (\bibinfo{year}{2012}).

\bibitem{Lebegue2010}
\bibinfo{author}{Leb\'{e}gue, ~S.} \emph{et~al.}
\newblock \bibinfo{title}{Cohesive properties and asymptotics of the dispersion
  interaction in graphite by the random phase approximation}.
\newblock \emph{\bibinfo{journal}{Phys. Rev. Lett.}}
  \textbf{\bibinfo{volume}{105}}, \bibinfo{pages}{196401}
  (\bibinfo{year}{2010}).

\bibitem{Liu2011}
\bibinfo{author}{Liu, Z.}, \bibinfo{author}{Liu, J.~Z.} \&
  \bibinfo{author}{et~al., Y.~C.}
\newblock \bibinfo{title}{Interlayer binding energy of graphite - a direct
  experimental determination}.
\newblock \emph{\bibinfo{journal}{arXiv}} \bibinfo{pages}{:1104.1469}
  (\bibinfo{year}{2011}).

\bibitem{Zacharia2004}
\bibinfo{author}{Zacharia, R.}, \bibinfo{author}{Ulbricht, H.} \&
  \bibinfo{author}{Hertel, T.}
\newblock \bibinfo{title}{van der Waals layered materials: opportunities and
  challenges}.
\newblock \emph{\bibinfo{journal}{Phys. Rev. B}}
  \textbf{\bibinfo{volume}{69}}, \bibinfo{pages}{155406}
  (\bibinfo{year}{2004}).

\bibitem{Graziano2012}
\bibinfo{author}{Graziano, G.}, \bibinfo{author}{Klime\v{s}, J.},
  \bibinfo{author}{Fernandez-Alonso, F.} \& \bibinfo{author}{Michaelides, A.}
\newblock \bibinfo{title}{Improved description of soft layered materials with
  van der Waals density functional theory}.
\newblock \emph{\bibinfo{journal}{Journal of Physics: Condensed Matter}}
  \textbf{\bibinfo{volume}{24}}, \bibinfo{pages}{424216}
  (\bibinfo{year}{2012}).

\bibitem{Wang2017}
\bibinfo{author}{Wang, L.} \emph{et~al.}
\newblock \bibinfo{title}{Superlubricity of a graphene/MoS$_2$ heterostructure: a
  combined experimental and dft study}.
\newblock \emph{\bibinfo{journal}{Nanoscale}} \textbf{\bibinfo{volume}{9}},
  \bibinfo{pages}{10846--10853} (\bibinfo{year}{2017}).

\bibitem{Rydberg2003}
\bibinfo{author}{Rydberg, H.} \emph{et~al.}
\newblock \bibinfo{title}{Van der Waals density functional for layered
  structures}.
\newblock \emph{\bibinfo{journal}{Phys. Rev. Lett.}}
  \textbf{\bibinfo{volume}{91}}, \bibinfo{pages}{126402}
  (\bibinfo{year}{2003}).

\bibitem{Benedict1998}
\bibinfo{author}{Benedict, L.~X.} \emph{et~al.}
\newblock \bibinfo{title}{Microscopic determination of the interlayer binding
  energy in graphite}.
\newblock \emph{\bibinfo{journal}{Chemical Physics Letters}}
  \textbf{\bibinfo{volume}{286}}, \bibinfo{pages}{490--496}
  (\bibinfo{year}{1998}).

\bibitem{Spanu2009}
\bibinfo{author}{Spanu, L.}, \bibinfo{author}{Sorella, S.} \&
  \bibinfo{author}{Galli, G.}
\newblock \bibinfo{title}{Nature and strength of interlayer binding in
  graphite}.
\newblock \emph{\bibinfo{journal}{Phys. Rev. Lett.}}
  \textbf{\bibinfo{volume}{103}}, \bibinfo{pages}{196401}
  (\bibinfo{year}{2010}).

\bibitem{Landolt2003}
\bibinfo{author}{Landolt-B\"{o}rnstein}.
\newblock \emph{\bibinfo{journal}{http://link.springer.de}}
  (\bibinfo{year}{2003}).

\bibitem{Isayev2017}
\bibinfo{author}{Isayev, O.}
\newblock \bibinfo{title}{Universal fragment descriptors for predicting
  properties of inorganic crystals}.
\newblock \emph{\bibinfo{journal}{Nat. Comm.}} \textbf{\bibinfo{volume}{8}},
  \bibinfo{pages}{15679} (\bibinfo{year}{2017}).

\bibitem{Tawfik2019}
\bibinfo{author}{Tawfik, S.~A.}, \bibinfo{author}{Isayev, O.},
  \bibinfo{author}{Stampfl, ~C.}, \bibinfo{author}{Shapter, ~J.}, \bibinfo{author}{Winkler, ~D. A.}, \&
  \bibinfo{author}{Ford, M.~J.}
\newblock \bibinfo{title}{Efficient prediction of structural and electronic
  properties of hybrid 2d materials using complementary DFT and machine
  learning approaches}.
\newblock \emph{\bibinfo{journal}{Advanced Theory and Simulations}}
  \textbf{\bibinfo{volume}{2}}, \bibinfo{pages}{1800128}
  (\bibinfo{year}{2019}).

\bibitem{Cai2018}
\bibinfo{author}{Cai, J.}, \bibinfo{author}{Luo, J.}, \bibinfo{author}{Wang,
  S.} \& \bibinfo{author}{Yang, S.}
\newblock \bibinfo{title}{Feature selection in machine learning:a new
  perspective}.
\newblock \emph{\bibinfo{journal}{Neurocomputing}}
  \textbf{\bibinfo{volume}{300}}, \bibinfo{pages}{70--79}
  (\bibinfo{year}{2018}).

\bibitem{Tamleh2018}
\bibinfo{author}{Tamleh, S.}, \bibinfo{author}{Rezaei, G.} \&
  \bibinfo{author}{Jalilian, J.}.
\newblock \bibinfo{title}{Stress and strain effects on the electronic structure
  and optical properties of scn monolayer}.
\newblock \emph{\bibinfo{journal}{Phys. Lett. A}}
  \textbf{\bibinfo{volume}{382}}, \bibinfo{pages}{339--345}
  (\bibinfo{year}{2018}).



\bibitem{Alexander2015}
\bibinfo{author}{Alexander, D.}, \bibinfo{author}{Tropsha, A.} \&
  \bibinfo{author}{Winkler, D.~A.}
\newblock \bibinfo{title}{Beware of R$^2$: correct statistical usage in qsar and
  qspr studies}.
\newblock \emph{\bibinfo{journal}{J. Chem. Inf. Model.}}
  \textbf{\bibinfo{volume}{55}}, \bibinfo{pages}{1316--1322}
  (\bibinfo{year}{2015}).

\bibitem{Materialsweb}
\bibinfo{title}{{Materialsweb}}.
\newblock \bibinfo{note}{Available at \url{https://materialsweb.org}}.

\bibitem{Sun2016}
\bibinfo{author}{Sun, W.} \emph{et~al.}
\newblock \bibinfo{title}{The thermodynamic scale of inorganic crystalline
  metastability}.
\newblock \emph{\bibinfo{journal}{Sci. Adv.}} \textbf{\bibinfo{volume}{2}},
  \bibinfo{pages}{1--8} (\bibinfo{year}{2016}).

\bibitem{WangNC2015}
\bibinfo{author}{Wang, W.} \emph{et~al.}
\newblock \bibinfo{title}{Measurement of the cleavage energy of graphite}.
\newblock \emph{\bibinfo{journal}{Nature Comm.}}
  \textbf{\bibinfo{volume}{6}} (\bibinfo{year}{2015}).

\bibitem{Ashton2016}
\bibinfo{author}{Ashton, M.}, \bibinfo{author}{Hennig, R.~G.},
  \bibinfo{author}{Broderick, S.~R.}, \bibinfo{author}{Rajan, K.} \&
  \bibinfo{author}{Sinnott, S.~B.}
\newblock \bibinfo{title}{Computational discovery of stable M$_2$AX phases}.
\newblock \emph{\bibinfo{journal}{Phys. Rev. B}} \textbf{\bibinfo{volume}{94}},
  \bibinfo{pages}{054116} (\bibinfo{year}{2016}).

\bibitem{Ashton2017}
\bibinfo{author}{Ashton, M.}, \bibinfo{author}{Paul, J.},
  \bibinfo{author}{Sinnott, S.~B.} \& \bibinfo{author}{Hennig, R.~G.}
\newblock \bibinfo{title}{Topology-scaling algorithm for bonded networks}.
\newblock \emph{\bibinfo{journal}{Phys. Rev. Lett.}}
  \textbf{\bibinfo{volume}{118}}, \bibinfo{pages}{106101}
  (\bibinfo{year}{2017}).

\bibitem{KresseJPCM1994}
\bibinfo{author}{Kresse, G.} \& \bibinfo{author}{Hafner, J.}
\newblock \bibinfo{title}{Norm-conserving and ultrasoft pseudopotentials for first-row and transition elements}.
\newblock \emph{\bibinfo{journal}{J. Phys.: Condens. Matt.}}
  \textbf{\bibinfo{volume}{6}}, \bibinfo{pages}{8245} (\bibinfo{year}{1994}).

\bibitem{KressePRB1999}
\bibinfo{author}{Kresse, G.} \& \bibinfo{author}{Joubert, D.}
\newblock \bibinfo{title}{From ultrasoft pseudopotentials to the projector augmented-wave method}.
\newblock \emph{\bibinfo{journal}{Phys. Rev. B}} \textbf{\bibinfo{volume}{59}},
  \bibinfo{pages}{1758} (\bibinfo{year}{1999}).



\bibitem{Tkatchenko2009}
\bibinfo{author}{Tkatchenko, A.} \& \bibinfo{author}{Scheffler, M.}
\newblock \bibinfo{title}{Accurate Molecular Van Der Waals Interactions from Ground-State Electron Density and Free-Atom Reference Data}.
\newblock \emph{\bibinfo{journal}{Phys. Rev. Lett.}}
  \textbf{\bibinfo{volume}{102}}, \bibinfo{pages}{073005}
  (\bibinfo{year}{2009}).

\bibitem{Tibshirani1996}
\bibinfo{author}{Tibshirani, R.}
\newblock \bibinfo{title}{Regression shrinkage and selection via the lasso}.
\newblock \emph{\bibinfo{journal}{Journal of the Royal Statistical Society}}
  \textbf{\bibinfo{volume}{58}}, \bibinfo{pages}{267--288}
  (\bibinfo{year}{1996}).

\bibitem{Forgy1965}
\bibinfo{author}{Forgy, E.~W.}
\newblock \bibinfo{title}{Cluster analysis of multivariate data: efficiency
  versus interpretability of classifications}.
\newblock \emph{\bibinfo{journal}{Biometrics}} \textbf{\bibinfo{volume}{21}},
  \bibinfo{pages}{768--769} (\bibinfo{year}{1965}).

\bibitem{Bergmann2014}
\bibinfo{author}{Bergmann, S.}, \bibinfo{author}{Stelzer, S.} \&
  \bibinfo{author}{Strassburger, S.}
\newblock \bibinfo{title}{On the use of artificial neural networks in
  simulation-based manufacturing control}.
\newblock \emph{\bibinfo{journal}{Journal of Simulation}}
  \textbf{\bibinfo{volume}{8}}, \bibinfo{pages}{76--90} (\bibinfo{year}{2014}).

\bibitem{MacKay1995}
\bibinfo{author}{MacKay, D. J.~C.}
\newblock \bibinfo{title}{Probable networks and plausible predictions - a
  review of practical bayesian methods for supervised neural networks}.
\newblock \emph{\bibinfo{journal}{Comput. Neural Sys.}}
  \textbf{\bibinfo{volume}{5}}, \bibinfo{pages}{469--505}
  (\bibinfo{year}{1995}).

\bibitem{MacKay1992}
\bibinfo{author}{MacKay, D. J.~C.}
\newblock \bibinfo{title}{A practical bayesian framework for backprop
  networks}.
\newblock \emph{\bibinfo{journal}{Neural Comput.}}
  \textbf{\bibinfo{volume}{4}}, \bibinfo{pages}{415--447}
  (\bibinfo{year}{1992}).

\bibitem{Burden1999}
\bibinfo{author}{Burden, F.~R.} \& \bibinfo{author}{Winkler, D.~A.}
\newblock \bibinfo{title}{Robust qsar models using bayesian regularized
  artificial neural networks}.
\newblock \emph{\bibinfo{journal}{J. Med. Chem.}}
  \textbf{\bibinfo{volume}{42}}, \bibinfo{pages}{3183--3187}
  (\bibinfo{year}{1999}).

\bibitem{Gal2016}
\bibinfo{author}{Gal, Y.} \& \bibinfo{author}{Ghahramani, Z.}
\newblock \bibinfo{title}{Dropout as a bayesian approximation: Representing
  model uncertainty in deep learning}.
\newblock \emph{\bibinfo{journal}{arXiv:1506.02142v6 [stat.ML]}}
  (\bibinfo{year}{2016}).

\bibitem{Srivastava2014}
\bibinfo{author}{Srivastava, N.}, \bibinfo{author}{Hinton, G.},
  \bibinfo{author}{Krizhevsky, A.}, \bibinfo{author}{Sutskever, I.} \&
  \bibinfo{author}{Salakhutdinov, R.}
\newblock \bibinfo{title}{Dropout: A simple way to prevent neural networks from
  overfitting}.
\newblock \emph{\bibinfo{journal}{Journal of Machine Learning Research}}
  \textbf{\bibinfo{volume}{15}}, \bibinfo{pages}{1929--1958}
  (\bibinfo{year}{2014}).

\bibitem{Tran2019}
\bibinfo{author}{Tran, D.}, \bibinfo{author}{Dusenberry, M.~W.},
  \bibinfo{author}{van~der Wilk, A.~M.} \& \bibinfo{author}{Hafner, D.}
\newblock \bibinfo{title}{Bayesian layers: A module for neural network
  uncertainty}.
\newblock \emph{\bibinfo{journal}{arXiv:1812.03973v3 [cs.LG]}}
  (\bibinfo{year}{2019}).

\end{thebibliography}
\end{document}